\documentclass[12pt]{article}
\usepackage{epsfig}
\usepackage{amsfonts}
\usepackage{amsmath}

\newcommand{\avg}[1]{\langle #1 \rangle}

\newcommand{\be}{\begin{equation}}
\newcommand{\ee}{\end{equation}}
\newcommand{\ba}{\begin{eqnarray}}
\newcommand{\ea}{\end{eqnarray}}
\newcommand{\bi}{\begin{itemize}}  
\newcommand{\ei}{\end{itemize}}

\newcommand{\R}{{\bf R}}

\newcommand{\Ocal}{{\mathcal O}}

\newcommand{\aslash}[1]{\,\,{\raise.15ex\hbox{/}\mkern-12mu #1}}
\newcommand{\bslash}[1]{\,\,{\raise.15ex\hbox{/}\mkern-9mu #1}}

\renewcommand{\bar}{\overline}
\renewcommand{\tilde}{\widetilde}

\newcommand\lrpar{\raise .8ex\hbox{$^\leftrightarrow$} \hspace{-9pt}
\partial}
\newcommand\lpar{\raise .8ex\hbox{$^\leftarrow$} \hspace{-9pt}
\partial}
\newcommand\rpar{\raise .8ex\hbox{$^\rightarrow$} \hspace{-9pt}
\partial}

\newcommand{\gsim}{\lower.7ex\hbox{$\;\stackrel{\textstyle>}{\sim}\;$}}
\newcommand{\lsim}{\lower.7ex\hbox{$\;\stackrel{\textstyle<}{\sim}\;$}}
\begin{document}

\baselineskip=18pt

\setcounter{footnote}{0}
\setcounter{figure}{0}
\setcounter{table}{0}

\begin{titlepage}

\begin{center}
\vspace{1cm}

{\Large \bf  More Holography from Conformal Field Theory}

\vspace{0.8cm}

{\bf Idse Heemskerk$^1$, James Sully$^1$}

\vspace{.5cm}

{\it $^1$ Department of Physics, University of California, \\ Santa Barbara, California
93106, USA}

\end{center}
\vspace{1cm}

\begin{abstract}
We extend the work of \cite{firstpaper} to support the conjecture that any conformal field theory with a large N expansion and a large gap in the spectrum of anomalous dimensions has a local bulk dual. We count to $O(1/N^2)$ the solutions to the crossing constraints in conformal field theory for a \textit{completely general} scalar four-point function and show that, to this order, the counting matches the number of independent interactions in a general scalar theory on Anti-de Sitter space. We introduce parity odd conformal blocks for this purpose.
\end{abstract}

\bigskip
\bigskip

%% \begin{flushleft}
%% June 2004
%% \end{flushleft}

\end{titlepage}

%%%%%%%%%%%%%%%%%%%%%%%%%%%%%%%%%%%%%%%%%%%%%%%%%%%%%%%%%%%%%%%
%\tableofcontents
%\vfill\eject
\section{Introduction}

Gauge/gravity duality \cite{juan}\cite{Witten}\cite{GKP} relates certain strongly coupled quantum field theories to weakly coupled theories of gravity, giving us perturbative access to the non-perturbative regime of these field theories. It is believed that any quantum field theory will have a stringy dual, but duality only has practical value if the string dual has a limit in which it reduces to local classical gravity where we can do perturbative calculations. It was recognized early on that a necessary condition for a field theory to have a local bulk dual like this is that it have a large $N$ limit as well as a hierarchy in the spectrum of anomalous dimensions. In a previous paper \cite{firstpaper}, we conjectured that this is in fact sufficient. To be precise,
\begin{quote}
Any CFT that has a planar expansion, and in which all single-trace operators of spin greater than two have parametrically large dimensions, has a local bulk dual.
\end{quote} 
For physical intuition and a discussion of the extent to which this had been tested in previous work, we refer to the introduction of \cite{firstpaper}. We provided evidence for this conjecture by studying the four point correlator of a single single-trace primary for a CFT of the conjectured form. To $O(1/N^2)$, and by restricting the spin, we counted the number of independent solutions to the crossing constraints for these correlators and showed that this matches the counting of bulk interaction Lagrangians, thereby confirming the conjecture. We did this in $d=2$ and $d=4$, making use of explicit expressions for conformal blocks. 
Matching the counting ruled out the logical possibility that there exist CFTs within this class corresponding to some smeared version of string theory on AdS or perhaps without an AdS description at all. Because the boundary correlator has the interpretation of a bulk S-matrix \cite{FSS}\cite{Gary:2009ae}, our result may also be seen as a proof in this setting of the lore that every S-matrix can be obtained from a local Lagrangian quantum field theory. 

In this paper we continue the previous work. We extend the CFT under consideration to contain a finite but arbitrary number of single-trace scalar operators and study the four-point correlator of four distinct operators. The counting changes in a non-trivial way, but is again shown to match, providing further evidence for the conjecture. In section \ref{cblocks} we rederive the explicit expressions for the conformal blocks. We  clarify the meaning of certain variables in which they take a simple form and we include the previously discarded possibility of parity odd conformal blocks in two dimensions. We explain the constraints imposed by crossing and clarify the perturbative expansion of the case of distinct degenerate operators where mixing occurs. In section \ref{generalscalar} we count solutions to crossing in CFT as well as bulk interactions and show that the numbers match. We discuss implications and possible future directions in section \ref{conclusion}.

\section{Conformal blocks and crossing}
\label{cblocks}

\subsection{Parity even and odd conformal blocks}
The four-point correlator in conformal field theory is naturally decomposed into a sum over conformal blocks; each block is the contribution of definite conformal Casimir, analogous to the partial wave decomposition of flat space amplitudes. A nice explicit form for the conformal blocks in $d=2,4$ was found by brute force in \cite{Dolan:2000ut} and subsequently in a more elegant way in \cite{Dolan:2003hv}. In this subsection we repeat the latter derivation, clarifying the meaning of the variables there introduced and including the omitted possibility of parity odd conformal blocks in $d=2$.

AdS$_{d+1}$---whose isometry group is the conformal group $O(d,2)$---can be represented by the surface $X^2 = -1$ in $d+2$ dimensional flat space with signature $--+\cdots +$. The boundary is then given by the projective null cone $X^2=0$ with $X^A \sim \lambda X^A$ and conformal fields of weight $\Delta$ on the boundary are homogeneous functions $\Ocal(\lambda X) = \lambda^{-\Delta}\Ocal(X)$. The conformal generators are
\begin{equation} \label{conformalgenerators}
 L_{AB} = X_A \frac{\partial}{\partial X^B} - X_B \frac{\partial}{\partial X^A} \, .
\end{equation}
Conformal invariance implies that the scalar four-point function on the boundary can be written in the form 
\begin{equation} \label{amplitude}
A \equiv \avg{\Ocal_1(X_1)\Ocal_2(X_2)\Ocal_3(X_3)\Ocal_4(X_4)} = \left( \frac{X_{24}}{X_{14}} \right)^{\Delta_1 - \Delta_2} \left( \frac{X_{14}}{X_{13}} \right)^{\Delta_3 - \Delta_4} \frac{\mathcal{A}_s(u,v)}{X_{12}^{\Delta_1+\Delta_2}X_{34}^{\Delta_3+\Delta_4}} 
,
\end{equation}
where $\mathcal{A}_s$ is the reduced amplitude which is a function of the conformally invariant cross ratios
\begin{equation}
u = \frac{X^2_{12}X^2_{34}}{X^2_{13}X^2_{24}}   \,\,\, , \,\,\,  v = \frac{X^2_{14}X^2_{23}}{X^2_{13}X^2_{24}}, \qquad X^2_{ij} = (X_i-X_j)^2
\, .
\end{equation}
We can break the amplitude up into contributions of fixed conformal Casimir in the 1-2 ($s$) channel $A= \sum_{C_{E,l}} p_{E,l} A_{E,l}$ with 
\begin{equation} \label{partialamplitudeDE}
 L_s^2 A_{E,l} = -C_{E,l} A_{E,l}, \qquad C_{E,l} = \frac{1}{2}\left(E(E-d) + l(l+d-2)\right) \, , \qquad L_s^2 = \frac{1}{4}(L_{1AB} + L_{2AB})^2 \, ,
\end{equation}
where the $A_{E,l}$ have to have the same form as the total amplitude
\begin{equation} \label{partialamplitude}
 A_{E,l} = \left( \frac{X_{24}}{X_{14}} \right)^{\Delta_1 - \Delta_2} \left( \frac{X_{14}}{X_{13}} \right)^{\Delta_3 - \Delta_4} \frac{g_{E,l}(u,v)}{X_{12}^{\Delta_1+\Delta_2}X_{34}^{\Delta_3+\Delta_4}} 
\end{equation}
The $g_{E,l}(u,v)$ are referred to as conformal partial waves and they have to obey a differential equation in $u,v$ obtained from substituting (\ref{partialamplitude}) into (\ref{partialamplitudeDE}).

In $d=2$ the connected part of the conformal group factorizes $SO(2,2) = SL_+(2,\R)\times SL_-(2,\R)$. This becomes manifest after a change of basis
\begin{equation} \label{newbasis}
L^\pm_x = \frac{1}{2}(L_{01} \pm L_{23}), \qquad L_y^\pm = \frac{1}{2}(L_{02}\mp L_{13}), \qquad L_z^\pm = \frac{1}{2}(J_{12}\pm J_{03}),
\end{equation}
and (dropping the label of the sub-algebra) we can subsequently change to a standard basis for the global subgroup of the classical Virasora algebra by
\begin{equation} \label{SL2Rgenerators}
 L_0 = - L_y, \qquad L_{\mp 1} = \pm(L_x \pm L_z)  \qquad \implies [L_m,L_n] = (m-n)L_{m+n} ,
\end{equation}
with Casimir
\begin{equation}
 L^2 = \frac{1}{2}(L_1L_{-1}+L_{-1}L_1) - L_0^2  \, .
\end{equation}
The Casimir of $SO(2,2)$ is now the sum of the two $SL(2,\R)$ Casimirs.
AdS$_{2+1}$ is invariant under $O(2,2)$ and we see from (\ref{conformalgenerators}) that the parity operation $X^A\to -X^A$ for any $A$  interchanges the $SL(2,\R)$.

Poincare coordinates on AdS$_{d+1}$ are given by \cite{BKL}
\begin{equation}
 X^{0} = \frac{r^2+1+\eta_{\mu\nu}y^\mu y^\nu}{2r}, \qquad X^d = -\frac{r^2-1+\eta_{\mu\nu}y^\mu y^\nu}{2r}, \qquad X^\mu = \frac{y^\mu}{r} \quad (\mu = 1\cdots d-1)
\end{equation}
with line element
\begin{equation}
 ds^2 = \frac{dr^2 + \eta_{\mu\nu}dy^\mu dy^\nu}{r^2}, \qquad \eta_{\mu\nu} = \mathrm{diag}(-,+ \cdots +) \, .
\end{equation}
In $d=2$ we change coordinates on the boundary to $x=y+t, \bar x=y-t$ and in terms of those the generators (\ref{SL2Rgenerators}) become 
\begin{equation} \label{PoincareSL2R}
 L_0 = -\frac{1}{2}r\partial_r - x\partial_x, \qquad L_{-1} = -\partial_x, \qquad L_{1}= -xr\partial_r - x^2\partial_x - r^2 \partial_{\bar x}
\end{equation}
and a similar set with $x\leftrightarrow \bar x$. After a Wick rotion $t\to it$ to Euclidean AdS $x$ and $\bar x$ are complex conjugates and acting on $r$ independent quantities at the surface $r=0$ the generators (\ref{PoincareSL2R}) reduce to the well known representation of the Virasora algebra on the plane $L_n = -x^{n+1}\partial$. To know how the AdS isometries act on boundary conformal fields of weight $\Delta$ we need the AdS/CFT dictionary
\begin{equation}
 (L_s^2)^+\avg{\Ocal_1(z_1,\bar z_1)\cdots \Ocal_4(z_4,\bar z_4)} = \lim_{r_i\to 0} \prod_i r_i^{-\Delta_i} (L_s^2)^+ \avg{\phi(x_1,r_1)\cdots\phi(x_4,r_4)} \, ,
\end{equation}
which means that when acting on boundary correlators we replace $r \partial_{r}$ by $\Delta$ in (\ref{PoincareSL2R}) and then take $r\to0$:
\begin{equation} \label{PoincareBdrySL2R}
 L_0 = -\frac{1}{2}\Delta - x\partial_x, \qquad L_{-1} = -\partial_x, \qquad L_{1}= -x\Delta - x^2\partial_x.
\end{equation}
The $s$-channel left Casimir becomes
\begin{equation}
 (L_s^2)^+ = \frac{\Delta_1+\Delta_2}{2}\left(1-\frac{\Delta_1+\Delta_2}{2}\right) +\Delta_1 (x_1-x_2)\partial_2 - \Delta_2 (x_1-x_2)\partial_1 
  + (x_1-x_2)^2\partial_1\partial_2.
\end{equation}
In terms of $x, \bar x$ the conformal cross ratios on the boundary can be written 
\begin{equation}
 u = z\bar z,\qquad  v=(1-z)(1-\bar z), \qquad z = \frac{x_{12}x_{34}}{x_{13}x_{24}}, \qquad x_{ij} = x_i - x_j.
\end{equation}
and the differential equation for the conformal partial waves obtained from substituting (\ref{partialamplitude}) in (\ref{partialamplitudeDE}) separates into left and right moving parts
\begin{align} \label{CPWDE}
 (D + \bar D)g_{E,l}(z,\bar z) &= \left(h(h-1)+\bar h(\bar h-1)\right)g_{E,l}(z,\bar z) =C_{E,l} g_{E,l}(z,\bar z)
\end{align}
where the left and right conformal weights are related to the spin and conformal dimension of the partial wave by $E = h+\bar h$, $l=h-\bar h$ and 
\begin{equation}
 D = z^2 \partial\left(1-\left(1+\Delta_{34}-\Delta_{12}\right)z\right)\partial + abz, 
\qquad \Delta_{ij} \equiv \frac{1}{2}(\Delta_i-\Delta_j) \, .
\end{equation}
The conformal partial waves therefore factorize $g_{E,l}(z,\bar z)=g_{h}(z)g_{\bar h}(\bar z)$, with 
\begin{equation}
 Dg_h = h(h-1)g_h  \implies g_h(z) = z^h \,_2F_1(h-\Delta_{12},h+\Delta_{34},2h;z) \, ,
\end{equation}
and we diagonalize parity by taking the linear combinations
\begin{equation} \label{parityoddandeven}
 g^\pm_{E,l}(z,\bar z) = g_h(z)g_{\bar h}(\bar z) \pm g_{\bar h}(z)g_{h}(\bar z) \qquad\qquad \textnormal{for } d=2
\end{equation}
with the plus (minus) corresponding to parity even (odd) CPWs. Whenever we drop the superscript in the rest of this paper we will be working with the parity even combination. 
In $d=4$ there are no parity odd conformal blocks because the conformal group doesn't factorize. Parity simply interchanges different states within the same representation (we can rotate out of the plane spanned by $z,\bar z$). A simple expression for the $d=4$ conformal blocks was derived in \cite{Dolan:2003hv}, but we will not present explicit equations for $d=4$ in this paper.

\subsection{Constraints from crossing}

CFT correlators are constrained not only by conformal symmetry, but also by crossing symmetry, which can be understood as associativity of the operator product. The operator product expansion (OPE) of two scalar operators in a CFT has the general form
\begin{equation}
\mathcal{O}_i(x) \mathcal{O}_j(0) = \sum x^{\Delta_k - \Delta_i - \Delta_j} c^k_{ij} \mathcal{O}_k(0)
\,  ,
\end{equation}
and is convergent whenever the distance $x$ between the two operators is less than distance to another operator. By taking the OPE of two pairs of nearby operators the four-point function reduces to a convergent expansion in a sum of two-point functions
\begin{equation}
\langle \mathcal{O}_i \mathcal{O}_j\mathcal{O}_k\mathcal{O}_l \rangle = \sum_{m} x_{ij}^{\Delta_m- \Delta_i - \Delta_j} c^{m}\!_{ij } \, x_{kl}^{\Delta_m - \Delta_k - \Delta_l}c_{mkl} \langle \mathcal{O}_m  \mathcal{O}_m \rangle
\, . \label{doubleope}
\end{equation}
There exist regions where this double OPE expansion is convergent for two different pairings of operators; we then must have that the amplitudes found using either of the expansions are equal:
\begin{equation}
\sum_{m} x_{ij}^{\Delta_m- \Delta_i - \Delta_j} c^{m}\!_{ij } \, x_{kl}^{\Delta_m - \Delta_k - \Delta_l}c^{m}\!_{kl} \langle \mathcal{O}_m  \mathcal{O}_m \rangle = \sum_{p} x_{ik}^{\Delta_p- \Delta_i - \Delta_k} c^{p}\!_{ik } \, x_{jl}^{\Delta_p - \Delta_j - \Delta_l}c^{p}\!_{jl} \langle \mathcal{O}_p  \mathcal{O}_p \rangle
\, . 
\end{equation}
This necessary equality is the crossing constraint. Note that while we could sum over the entire set of operators in the theory on either side, we generically expect that different subsets of the operators will have non-vanishing coefficients on either side of the crossing equation.

We can group the operators contributing to the OPE into representations of the conformal group consisting of primary operators $\mathcal{O}_P$ and their descendents (from acting on $\mathcal{O}_P$ with derivatives). We can then write the four-point function in terms of these conformal blocks
\begin{eqnarray}
\langle \mathcal{O}_i \mathcal{O}_j\mathcal{O}_k\mathcal{O}_l \rangle &=& \sum_{P} c^{P}\!_{ij } c_{Pkl} \langle (x_{ij}^{\Delta_P- \Delta_i - \Delta_j}\mathcal{O}_P + ... )(  \, x_{kl}^{\Delta_P - \Delta_k - \Delta_l}\mathcal{O}_P + ...) \rangle \nonumber\\
 &=& \sum_{P} c^{P}\!_{ij } c_{Pkl} \mathbb{CB}_P(x_{ij},x_{kl})
\, . \label{blockope}
\end{eqnarray}
The conformal blocks transform in the same way as the full correlator and can therefore be written as a prefactor times a conformally invariant reduced part that depends only on the cross ratios $z,\bar z$. The reduced conformal blocks are exactly the conformal partial waves from the previous section. We can identify the coefficients of the CPWs and the coefficients of the OPE expansion as
\begin{equation}
 p(h_P,\bar{h}_P)  =  \sum_{P} c^{P}\!_{ij } c^{P}\!_{kl}
 \, .
\end{equation}

From the definition of $z,\bar z$ we see that the limit $x_1\to x_2$ corresponds to the limit $z\to 0$. Likewise, the limit $x_1\to x_4$ corresponds to $1-z\to 0$, and $x_1\to x_3$ corresponds to $1/z\to 0$. We identify these limits as the $s, \, t$ and $u$ channels respectively. Expanding the scattering amplitude \eqref{amplitude} in appropriate CPWs to these limits corresponds to taking a double-OPE and grouping into conformal blocks. Equating the expansions is just a restatement of associativity of the OPE. When we equate the expansion in two channels, the prefactors must necessarily combine into a conformally invariant function of $z, \bar z$. We find, in particular,
\begin{eqnarray}
[z\bar{z}]^{\frac{-\Delta_1 - \Delta_2}{2}} \mathcal{A}_s(z,\bar{z}) &=&  [(1-z)(1-\bar{z})]^{\frac{-\Delta_2 - \Delta_3}{2}} \mathcal{A}_t(1-z,1-\bar{z}) \nonumber \\
\mathcal{A}_s(z,\bar{z}) &=& [z\bar{z}]^{\frac{\Delta_1+ \Delta_4}{2}}\mathcal{A}_u({1}/{z},{1}/{\bar{z}})
\, .
\label{crossing}
\end{eqnarray}
The overlapping regions of convergence of the OPE mean that these expansions can be directly compared.\footnote{Although the overlapping regions of convergence make this story clear, they will not be of primary importance in our derivations. In fact, the identities of hypergeometric functions will allow us to analytically continue from one channel completely over to the domain of convergence of another.}

% This is the form of the crossing constraint that will be directly used in the rest of this paper.
% 
% Lastly, we note that the expansion in terms of conformal partial waves described Section 2.1
% \begin{equation}
% \mathcal{A} = \sum_{h,\bar{h}} p(h,\bar{h}) g_{h,\bar{h}}
% \, .
% \end{equation}
% is equivalent to the expansion of the double-OPE in terms of conformal blocks. We make the easy identification
% \begin{eqnarray}
% p(h_P,\bar{h}_P)  &=&  c^{P}\!_{ij } c_{Pkl} \\
% \left( \frac{X_{24}}{X_{14}} \right)^{\Delta_1 - \Delta_2} \left( \frac{X_{14}}{X_{13}} \right)^{\Delta_3 - \Delta_4} \frac{g_{h_P,\bar{h}_P}}{X_{12}^{\Delta_1+\Delta_2}X_{34}^{\Delta_3+\Delta_4}} &=& \mathbb{CB}_P
% \, .
% \end{eqnarray}

\subsubsection{$1/N$ Expansion}

We will solve the crossing relations in the $1/N$ expansion. We write this expansion in the form
\begin{eqnarray}
{\mathcal A}(z,\bar{z})  &=& {\mathcal A}_0(z,\bar{z})  + \frac{1}{N^2}{\mathcal A}_1(z,\bar{z})  + \ldots\ , \nonumber\\
p(h,\bar{h}) &=& p_0(h,\bar{h}) + \frac{1}{N^2} p_1(h,\bar{h}) + \ldots\ , \nonumber\\
E(h,\bar{h}) &=& h_0+\bar{h}_0 +  \frac{1}{N^2} \gamma_1(h,\bar{h}) + \ldots \ .
\end{eqnarray}
Thus at zeroth order in $1/N^2$ we have
\begin{eqnarray}
{\mathcal A}_0(z,\bar z)  =  \sum_{h,\bar{h}} p_0(n,l) \,{g_{h,\bar{h}} (z,\bar{z})} \ ,
\end{eqnarray}
and at first order
\begin{eqnarray}
{\mathcal A}_1(z,\bar z)  = 
\sum_{h,\bar{h}} p_1(h,\bar{h}) \,{g_{h_0,\bar{h}_0} (z,\bar{z})}
+  p_0(h,\bar{h}) \gamma_1(h,\bar{h}) \, \frac{\partial}{\partial E} {g_{h_0,\bar{h}_0} (z,\bar{z})}
\, .
\label{A1cross}
\end{eqnarray}
Because some operators may have degenerate dimension at $0$-th order, they will have the same CPWs $g_{h_0,\bar{h}_0}$ at this order in the expansion. The separate contributions of these operators will not be distinguishable in our analysis at this order. This is discussed in more detail in the following section.

\subsubsection{Degeneracy}
\label{degeneracysection}
One would expect from (\ref{A1cross}) that at $O(1/N)$ there are no contributions to the amplitude from anomalous dimensions of operators whose OPE coefficient is $O(1/N)$, or in other words, from operators that have $p_0=0$. However, as was explained in \cite{D'Hoker:1999jp}, operators with degenerate bare dimension are mixed by interactions and therefore this naive expectation is incorrect.

Suppose we start with an orthonormal set of scalar primaries of bare dimension $E^{(0)}$, some of which may be degenerate
\begin{equation} \label{ortho}
 \avg{\Ocal_{\alpha}(x)\Ocal_\beta(0)}=\frac{\delta_{\alpha\beta}}{|x|^{2E^{(0)}}}.
\end{equation}
When we include interactions, this degeneracy is lifted and our original orthonormal basis may not coincide with the non-degenerate eigenstates of the interacting dilation operator.
\begin{equation}
 [D,\Ocal_\alpha] = M_{\alpha\beta} \Ocal_\beta = (\delta_{\alpha\beta}E^{(0)}_{\alpha} + \gamma_{\alpha\beta})\Ocal_\beta
\end{equation}
We therefore do a basis transformation $\gamma_{\alpha\beta} V^{\alpha}_{\,\,A} V^\beta_{\,\,B} =  \textnormal{diagonal}$ and get
\begin{equation}
 \avg{\Ocal_A(x) \Ocal_B(0)} = \delta_{AB} \avg{\Ocal_A(x) \Ocal_A(0)} \implies  \avg{\Ocal_{\alpha}(x)\Ocal_\beta(0)} = V_{\alpha}^{\,\,A} V_\beta^{\,\,B} \frac{\delta_{AB}}{|x|^{2\Delta+2\gamma_A}}
\end{equation}
It is the $\Ocal_A$ that transform in representations of the conformal group at $O(1/N^2)$ and that therefore correspond to the partial waves
\begin{align} \label{diagonalOPE}
 A&= \avg{\Ocal_1(x_1)\Ocal_2(x_2)\Ocal_3(x_3)\Ocal_4(x_4)}=  \sum_A c_{12\Ocal_A}c_{34 \Ocal_A} \mathbb{CB}_{A}(x_i) \equiv  \sum_A p_A \mathbb{CB}_{A}(x_i)
  \\\nonumber
 &= \sum_{E^{(0)},l}\sum_i \left(p^{(0)}_{i}(E^{(0)},l) + p^{(1)}_{i}(E^{(0)},l) + p^{(0)}_{i}(E^{(0)},l) \gamma_{i}(E^{(0)},l) \frac{\partial}{\partial E}\right)\mathbb{CB}_{i}(x_i)
\end{align}
% Note that $\alpha,\beta$ do not run over the same set, they denote operators that can appear in the OPEs $\Ocal_1\Ocal_2, \Ocal_3\Ocal_4$ respectively. 
where in the second line we have split the sum over $A$ into one over subspaces of different bare dimension and a sum within barely degenerate subspaces. Parenthesised superscripts denote order in the perturbation expansion. It is the outer sum in the second line that runs over distinct partial waves and so we identify
\begin{align}
 p^{(0)}(E^{(0)},l) &\equiv \sum_i p^{(0)}_{i}(E^{(0)},l), \qquad p^{(1)}(E^{(0)},l) \equiv \sum_i p^{(1)}_{i}(E^{(0)},l)
\\\nonumber
p^{(0)}(E^{(0)},l)\gamma(E^{(0)},l) &\equiv \sum_i p^{(0)}_{i}(E^{(0)},l) \gamma_{i}(E^{(0)},l)
\end{align}
$p^{(0)}$ is related to the zeroth order OPE coefficients in our original basis \eqref{ortho} (that we used to express our external states in and which was a good basis at zeroth order) by
\begin{equation}
 p^{(0)}(E^{(0)},l) = \sum_i \sum_{\alpha,\beta}c^{(0)}_{12\alpha} c^{(0)}_{34\beta}V_{\alpha}^{\,\,i} V_{\beta}^{\,\,i} 
\end{equation}
Now comes an important point. At zeroth order there will be never be degenerate operators appearing in the same OPE, for example, $\Ocal_1(0)\Ocal_2(x) \sim \Ocal_1\Ocal_2(0)+\cdots$ with $\Ocal_1$, $\Ocal_2$ of dimension $\Delta$ contains only one operator of bare dimension $2\Delta$, namely $\Ocal_{\{12\}}(x) \equiv \Ocal_1\Ocal_2(x)$. 
Therefore the zeroth order OPE coefficients are like Kronecker deltas 
\begin{equation}
 c^{(0)}_{\alpha\beta\{\gamma\delta\}} \sim \delta^\alpha_{(\gamma}\delta^\beta_{\delta)}
\end{equation}
and the sum over $\alpha,\beta$ can be dropped in $p^{(0)}(E^{(0)},l)$. The sum over $i$ then gives $p^{(0)}(E^{(0)},l)=0$, as expected since the zeroth order contribution to the amplitude must vanish, and we see that
\begin{align}\label{anomalousdimensions}
 \gamma(E^{(0)},l) &= \sum_i V_{\{12\}}^{\,\,i} V_{\{34\}}^{\,\,i} \gamma_{i}(E^{(0)},l) = \gamma_{\{12\} \{34\}},
\qquad p_0(E^{(0)},l) = c^{(0)}_{12\{12\}} c^{(0)}_{34\{34\}} \, .
\end{align}
Note that the anomalous dimension just becomes the off-diagonal component of the interacting dilation operator. The first order part of the amplitude is now
\begin{align}
 A^{(1)}
= \sum_{E^{(0)},l} \Big(p_{0}(E^{(0)},l) \gamma(E^{(0)},l) \partial_{E^{(0)}} + p_{1}(E^{(0)},l) \Big) \mathbb{CB}_{E^{(0)},l}(x_j)
\end{align}
This is of the same form as \eqref{A1cross} but we see that with our new definition of $p_0,\gamma$, neither is necessarily vanishing.

\subsubsection{The degenerate limit of a generic expansion}

We consider now an expansion where there are no degenerate bare dimensions. We then let two dimensions grow arbitratily close, or more generally, two sets of operators grow pairwise close. We demonstrate how the anomalous dimensions in the degenerate case can then be extracted from the coefficients of CPWs in the generic case (at first order in $1/N^2$).

Consider such a pair of operators, $\mathcal{O}_{\{ 12 \}}$ of dimension $\Delta$ and $\mathcal{O}_{\{ 34 \}}$ of dimension $\Delta +\epsilon$. We have the interacting dilation operator
\begin{equation}
 [D,\Ocal_\alpha] = M_{\alpha\beta} \Ocal_\beta = (\text{Diag}(\Delta,\Delta + \epsilon)_{\alpha\beta} + \frac{1}{N^2} \gamma_{\alpha\beta})\Ocal_\beta \, .
\end{equation}
Analogously to non-degenerate quantum mechanical perturbation theory, it is easy to see that the transformation that diagonalizes dilation is given to leading order in $1/N^2$ by
\begin{equation}
V_{\alpha\beta} = \delta_{\alpha\beta} + \epsilon_{\alpha\beta} \frac{\gamma_{\alpha\beta}}{\epsilon N^2}  
\end{equation}
Labeling the new eigenstates by their zeroth order state, we have find their coefficient in the scattering amplitude is given by
\begin{eqnarray}
 p^{(1)}_{\{ 12 \}}  &=& c^{(0)}_{ 12 \{ 12 \}} c^{(0)}_{34 \{ 34 \}} \left( \frac{\gamma_{\{ 12 \}\,\,34}}{\epsilon N^2}\right) \nonumber \\
 p^{(1)}_{\{ 34 \}}  &=& c^{(0)}_{ 12 \{ 12 \}} c^{(0)}_{34 \{ 34 \}} \left(- \frac{\gamma_{\{ 34 \}\,\,\{ 12 \}}}{\epsilon N^2} \right) \, .
\end{eqnarray}
It is then immediate, comparing to \eqref{anomalousdimensions}, that we can extract the anomalous dimension from the coupling as
\begin{equation}
p_0(E^{(0)},l) \gamma(E^{(0)},l) = \lim_{\epsilon \rightarrow 0} \pm \epsilon p^{(1)}_{\alpha} 
\end{equation}
where $\alpha$ denotes one of the two operators that becomes degenerate with the operator with dimension $E^{(0)}$ and the plus or minus sign is chosen accordingly. 

From this equality we can see that, for any generic amplitude, the first order coefficients give the anomalous dimensions of solutions in the degenerate limit of that amplitude, and every such degenerate amplitude can be found in this way by a small perturbation in the bare scaling dimensions. Thus the generic case gives an upper-bound on allowed amplitudes for the degenerate amplitude. Nevertheless, we will see in the following section that it will be most instructive to consider the degenerate case first. 

\section{Solving the general scalar model}
\label{generalscalar}
A CFT is completely specified by its OPE. If we specify the allowed operators in the OPE, crossing will give us a constraint on their OPE coefficients and the anomalous dimensions. The set of solutions to these constraints is the set of all CFTs with the specified operator content. In our previous paper \cite{firstpaper} we counted solutions to crossing for the four point function of a scalar single-trace operator $\Ocal$ having in its OPE all the double trace primaries that could be constructed from it, as well as any number of single trace operators such as the stress tensor $T_{\mu\nu}$. The non-trivial part of the counting concerned itself solely with the double traces while every single trace operator added to the theory gave only a single new solution to crossing.

We now extend our story to consider the most general scalar four-point correlator, which is of four different single trace operators. Further extensions would have to involve either higher spin external states, for which explicit expressions for the conformal blocks are currently unknown, or go to higher order in $1/N$, which is dual to loops in the bulk so that it is not clear that there would be finite solutions at all.

The distinct external operators lead to new double-trace operators appearing in the OPE and hence in the CPW expansion. Moreover, different double-trace operators will be exchanged in each channel. This extension is therefore non-trivial. As we will see, the number of solutions to crossing still matches the number of bulk interactions, providing further evidence for the conjecture that the class of CFTs under consideration is in one-to-one correspondence with local supergravity-type theories. 

\subsection{Crossing constraints: degenerate case}
\label{degenerateconstraintssection}
We begin by taking the external operators distinguishable, but with degenerate scaling dimension. Conveniently, the crossing equations are much simpler in this case, but the number of solutions remains the same as for the non-degenerate case. From the bulk perspective this is obvious because the number of possible interactions in the Lagrangian does not depend on the masses of the fields.
In this case, the spectrum of double trace operators is given by
\begin{equation}
{\cal O}_i \lrpar_{\mu_1} \ldots  \lrpar_{\mu_l}  (  \lrpar_{\nu} \,  \lrpar^{\nu })^n {\cal O}_j - {\rm traces} \ ,
\label{double}
\end{equation}
with bare dimension $E=2\Delta + 2n +l$. 
In $d=2$ these are reducible and the irreducible parts are the components with all indices $z$ or all indices $\bar z$ (other components vanish), which are interchanged by parity. Therefore, it is not natural to work with the parity even or odd combinations of these two representations \eqref{parityoddandeven}, which may get different anomalous dimensions in a parity violating theory. However, at least algebraically, we can always split the amplitude into even and odd parts under $z \leftrightarrow \bar z$ and these will be equal to the respective even and odd parts on the other side of the crossing constraint. We can therefore count solutions using \eqref{parityoddandeven}. This is natural from the bulk perspective because at leading order in $O(1/N^2)$ we expect a one-to-one correspondence between even (odd) solutions to crossing and even (odd) interaction terms in the Lagrangian.
From (\ref{crossing}) and (\ref{A1cross}), the parity even crossing constraints are then
\begin{align} \label{degenerateunprojectedcrossing}
\frac{1}{(z\bar z)^{\Delta}} & \sum_n \sum_{l=0}^{L_s} \left(p_1^{(s)}(n,l) + p^{(s)}_0(n,l)\frac{1}{2} \gamma^{(s)}_{n,l} \partial_n  \right)  g_{2\Delta +2n +l,l}(z,\bar z)
\\\nonumber 
= \frac{1}{((1-z)(1-\bar z))^{\Delta}} &\sum_n \sum_{l=0}^{L_t}\left( p_1^{(t)}(n,l) + p_0^{(t)}(n,l)\frac{1}{2} \gamma^{(t)}_{n,l} \partial_n \right) g_{2\Delta +2n +l,l}(1-z,1-\bar z)
\\\nonumber
= & \sum_n \sum_{l=0}^{L_u} \left(p_1^{(u)}(n,l) + p^{(u)}_0(n,l)\frac{1}{2} \gamma^{(u)}_{n,l}\partial_n  \right) g_{2\Delta +2n +l,l}(1/z,1/\bar z) .
\end{align}
We have restricted the spin in all channels to get a finite number of solutions. Using hypergeometric identities we can analytically continue all three of these to relations between expansions around $z=0$ and $z=1$. For example, $tu$ crossing becomes
\begin{align}
\frac{1}{(z\bar z)^{\Delta}} &\sum_n \sum_{l=0}^{L_t}\left( p_1^{(t)}(n,l) + p_0^{(t)}(n,l)\frac{1}{2} \gamma^{(t)}_{n,l} \partial_n \right) (-1)^l g_{2\Delta +2n +l,l}\left(z,\bar z\right)
\\\nonumber
= &\frac{1}{((1-z)(1-\bar z))^{\Delta}}\sum_n \sum_{l=0}^{L_u} \left(p_1^{(u)}(n,l) + p^{(u)}_0(n,l)\frac{1}{2} \gamma^{(u)}_{n,l} \partial_n  \right) g_{2\Delta +2n +l,l}(1-z,1-\bar z) 
\end{align}
To obtain equations that involve only $\gamma$, and to remove dependence on $\partial_n$ and $p_1$, we take the part proportional to $\ln(1-\bar z)\ln z$. We throw away no information by considering only this part because the parts of the equation that are proportional to a log in just one of the variables determine the $p_1$ in terms of the $\gamma$.
The $\ln z$ comes from the hypergeometric identity
\begin{equation}
 F_a(1-z) = \ln z \tilde F_a(z) + \textnormal{holomorphic at } z=0
\end{equation}
on one side, and from the $\partial_n$ on the other side. Here we have defined
\begin{equation} \label{Fdef}
 F_a(z) \equiv \,_2 F_1 (a,a;2a;z), \qquad \tilde F_a(z) \equiv \,_2 F_1 (a,a;1;z) \, .
\end{equation}
Then using the orthogonality relation
\begin{equation} \label{degenerateorthogonality}
 \oint_C \frac{dz}{2\pi i} z^{m-m'-1}F_{\Delta+m}(z) F_{1-\Delta-m'}(z) = \delta_{mm'},
\end{equation}
with $C$ a contour around the origin, we project out terms of fixed $n$ around $z=0$ and $\bar z=1$ and obtain
\begin{align}\label{degeneratecrossing}
\sum_{l=0}^{L_s} \left[\gamma^{(s)}_{p,l} J(p+l,q) + \gamma^{(s)}_{p-l,l} J(p-l,q)\right]
&= \sum_{l=0}^{L_t} \left[ \gamma^{(t)}_{q,l} J(q+l,p) + \gamma^{(t)}_{q-l,l} J(q-l,p) \right]
\\\nonumber
\sum_{l=0}^{L_t}(-1)^l\left[ \gamma^{(t)}_{p,l} J(p+l,q) + \gamma^{(t)}_{p-l,l} J(p-l,q) \right]
&= \sum_{l=0}^{L_u} \left[ \gamma^{(u)}_{q,l} J(q+l,p) + \gamma^{(u)}_{q-l,l} J(q-l,p)\right]
\\\nonumber
\sum_{l=0}^{L_s}  (-1)^l \left[ \gamma^{(s)}_{p,l} J(p+l,q) + \gamma^{(s)}_{p-l,l} J(p-l,q) \right]
&= \sum_{l=0}^{L_u}  (-1)^l \left[ \gamma^{(u)}_{q,l} J(q+l,p) + \gamma^{(u)}_{q-l,l} J(q-l,p)\right].
\end{align}
Here we have absorbed $p_0^{(i)}$ into the definition of $\gamma^{(i)}$ and have defined a coefficient function
\begin{equation}
J(m,m') \equiv  \oint_C \frac{dz}{2\pi i} \frac{(1-z)^m}{z^{m'+1}} \tilde F_{\Delta+m}(z)F_{1-\Delta-m'}(z).
\end{equation}
Equation (\ref{degeneratecrossing}) and every equation from here on refers to $d=2$. For $d=4$ things work out in exactly the same way (we showed this for the simpler case in our previous paper) and is straightforward to obtain the analogous equations so we will not include them here.

Our goal is to count the number of solutions to (\ref{degeneratecrossing}). Without loss of generality we can take $L_s \leq L_t \leq L_u$ and we see immediately that we can have at most $(L_t+1)(L_s+1)$ solutions. This is because for fixed $q$ there are $L_t+1$ unknown $\gamma^{(t)}$ and if we specify the $\gamma^{(s)}(p,l)$ for $p\leq L_t$ then we have $L_t+1$ equations to solve for the unknown $\gamma^{(t)}$. We expect the actual number of solutions to be smaller because after having fixed the full amplitude with the $p\leq L_t$ equations, the $p>L_t$ equations provide further constraints on the specified block of $\gamma^{(s)}$. For example, if we take $p=L_t+1$ the RHS of the $st$-constraint involves the $L_s+1$ new variables $\gamma^{(s)}_{L_t+1,l}$ but also the $\gamma^{(s)}$ with $\max(0,L_t + 1-L_s)\leq p \leq L_t$ from the second term and there is a constraint on these variables for every $q$. For $p>L_s + L_t$ the equations no longer contain $\gamma^{(s)}$ from the initially specified block so we need not consider those. The number of solutions will be thus be reduced from $(L_s+1)(L_t+1)$ by the number of independent constraints with $L_t > p \geq L_t+L_s$, $q$ arbitrary. A similar argument holds for the other two constraints.

Because $q$ is arbitrary we have an infinite number of constraints on a finite number of variables so almost all of the constraints must be redundant if we are to have solutions at all. This is challenging to show and, given the facts established above, we will simply consider the $st$ and $tu$ equations with $p,q\leq 2L_u$ and let mathematica determine how many of those are independent. We found experimentally that further increasing the limit on $q$ and $p$ does not reduce the number of solutions and also that the $su$ equation provides no independent constraints\footnote{This is obvious when considering the full crossing equations but perhaps surprising for the equations with cut-off $p,q$.} so all the higher $q$ equations do indeed seem to be redundent. Nevertheless, we can keep open the possibility that the number of solutions does decrease by including higher $q$ and temporarily consider the number found this way an upper bound on the number of solutions. We will soon show that this upper bound saturates a lower bound derived from the bulk. The upper bound is given by
\begin{equation} \label{Jamiesform}
 (L_t+1)(L_s+1) - \frac{1}{2}(L_s +L_t-L_u)(L_s+L_t-L_u+1) + \lfloor(L_s+L_t-L_u)^2/4\rfloor
 \, ,
\end{equation}
where $\lfloor ... \rfloor$ denotes the floor operator.

For parity odd intermediate states in $d=2$ we find the same expression with $L_i\to L_i-1$. This is intuitive because the parity odd conformal blocks vanish for $l=0$ which makes all the counting start at $l=1$. In $d=4$ there are no parity odd four point amplitudes and therefore, as explained in section 2, there are no parity odd conformal blocks.
Bulk counting in the next section will provide a lower bound on the number of solutions and we will find that this matches the upper bound (\ref{Jamiesform}), demonstrating a one to one correspondence between local theories in the bulk and boundary CFTs.

\subsection{Crossing constraints: generic general case}

We now consider the generic general case where the scattered scalars have completely generic scaling dimensions. In particular, we will assume that the scalars have incommensurate non-integer dimensions. As will be clear, this allows a particular analytic continuation of the CPWs and allows a useful differentiation between branch cuts which will be used in solving the crossing constraints. 

In the $s$-channel expansion, the contributing CPWs to the four-point function will be exactly those corresponding to double-trace operators of $\mathcal{O}_1\mathcal{O}_2$ and $\mathcal{O}_3\mathcal{O}_4$ with bare scaling dimensions $E_{12}(n,l) = \Delta_1 + \Delta_2 + 2n + l$ and $E_{34}(n,l) = \Delta_3 + \Delta_4 + 2n + l$. Likewise, in the t-channel, we expand in double-trace formed from  $\mathcal{O}_2\mathcal{O}_3$ and  $\mathcal{O}_1\mathcal{O}_4$ and in the u-channel from  $\mathcal{O}_1\mathcal{O}_3$ and  $\mathcal{O}_2\mathcal{O}_4$. 

Because the four scattered scalars are distinct, the disconnected contribution to the scattering amplitude will vanish and there will be no contribution at order $(\tfrac{1}{N})^0$. The crossing constraints take the form
\begin{align} \label{genericunprojectedcrossing}
\frac{1}{(z\bar z)^{\tfrac{\Delta_1 + \Delta_2}{2}}} & \sum_{\substack{E \, = \, E_{12}(n,l) \,, \\ E_{34}(n,l)}} \sum_{l=0}^{L_s} p_1^{(s)}(E,l)  g_{E,l}(z,\bar z)
\\\nonumber 
= \frac{1}{((1-z)(1-\bar z))^{\tfrac{\Delta_2 + \Delta_3}{2}}} &\sum_{\substack{E \, = \, E_{23}(n,l) \,, \\ E_{14}(n,l)}} \sum_{l=0}^{L_t} p_1^{(t)}(E,l) g_{E,l}(1-z,1-\bar z)
\\\nonumber
= \frac{1}{(z\bar z)^{\tfrac{\Delta_2 - \Delta_4}{2}}} & \sum_{\substack{E \, = \, E_{13}(n,l) \,, \\ E_{24}(n,l)}} \sum_{l=0}^{L_u} p_1^{(u)}(E,l)  g_{E,l}(1/z,1/\bar z) .
\end{align}
Note that despite the compressed notation, we sum over two distinct towers of conformal partial waves in each channel (eg. $E_{12}(n,l)$ and $E_{34}(n,l)$ in the s-channel) and so in the general case the coefficients $p(E_{ij},l)$ are indexed by the specific double-trace operators as well as by $n$ and $l$. The crossing equation \eqref{genericunprojectedcrossing} also differs from the degenerate case \eqref{degenerateunprojectedcrossing} in that there are no anomalous dimensions contributing to this order.

We solve the crossing constraints by comparing the $st$ and $su$ channel equations. The $ut$ channel is redundant as discussed above. First we analytically continue the hypergeometric functions in $z$ or $\bar z$ in each channel to the appropriate region of convergence in the other channel. We do this using the identities for the analytic continuation of hypergeometric functions with generic arguments, listed in Appendix \ref{identities}. The analytic continuation generates terms with two different branch cuts for each tower of double-trace operators. In every case, the two branch cuts corresponding to a single double-trace tower in one channel match exactly one of the two branch cuts for each tower in the opposite channel. Thus, by looking at terms with a specific branch structure in the crossing equations we can constrain all of the coefficients by specifying sufficient free coefficients for one tower. As an example, the $st$ branch cuts are listed in Table \ref{branchcuts}; the other channels can be quickly computed to show the same structure.

\begin{table}
\centering
\begin{tabular}{|l||c|c|}
\hline
$p^{(s)}(E_{12},l)$ & $(1-z)(\bar{z})$ &  $(1-z)^{\tfrac{\Delta_1 - \Delta_2 - \Delta_3 + \Delta_4 }{2}}(\bar{z})$  \\ \hline
$p^{(s)}(E_{34},l)$ & $(1-z)(\bar{z})^{\tfrac{-\Delta_1 - \Delta_2 + \Delta_3 + \Delta_4 }{2}}$ &  $(1-z)^{\tfrac{\Delta_1 - \Delta_2 - \Delta_3 + \Delta_4 }{2}}(\bar{z})^{\tfrac{-\Delta_1 - \Delta_2 + \Delta_3 + \Delta_4 }{2}}$  \\ \hline
$p^{(t)}(E_{23},l)$ & $(1-z)(\bar{z})^{\tfrac{-\Delta_1 - \Delta_2 + \Delta_3 + \Delta_4 }{2}}$ &  $(1-z)(\bar{z})$  \\ \hline
$p^{(t)}(E_{14},l)$ & $(1-z)^{\tfrac{\Delta_1 - \Delta_2 - \Delta_3 + \Delta_4 }{2}}(\bar{z})$ &  $(1-z)^{\tfrac{\Delta_1 - \Delta_2 - \Delta_3 + \Delta_4 }{2}}(\bar{z})^{\tfrac{-\Delta_1 - \Delta_2 + \Delta_3 + \Delta_4 }{2}}$  \\
\hline
\end{tabular}
\caption{The types of branch cuts found for each tower of double-trace operators in the $st$-crossing equations.}
\label{branchcuts}
\end{table}

Once we have isolated the terms with a specific branch cut, analogously to the degenerate case, we can construct projection operators for the relevant hypergeometric functions. The projection operators are constructed in Appendix \ref{projops} and are straightforward, albeit messy, generalizations of those used in the degenerate case. 

We project onto terms with fixed energies and spin in each channel to obtain crossing equations, exactly as in the degenerate case. We list below the $st$ and $su$ crossing equations where we have isolated branch cuts to constrain the  $\mathcal{O}_1\mathcal{O}_2$ and $\mathcal{O}_2\mathcal{O}_3$ towers in the $st$ relation and the $\mathcal{O}_1\mathcal{O}_2$ and the $\mathcal{O}_1\mathcal{O}_3$ towers in the $su$ relation:
\begin{align}\label{genericXing}
\sum_{l=0}^{L_s} \left[p^{(12)}_{p,l} J^{(st)}_{1,2,3,4}(p+l,q) + p^{(12)}_{p-l,l} J^{(st)}_{1,2,3,4}(p-l,q)\right]
&= \sum_{l=0}^{L_t} \left[ p^{(23)}_{q,l} J^{(st)}_{3,2,1,4}(q+l,p) + p^{(23)}_{q-l,l} J^{(st)}_{3,2,1,4}(q-l,p) \right]
\\\nonumber
\sum_{l=0}^{L_s} \left[p^{(12)}_{p,l} J^{(su)}_{1,2,3,4}(p+l,q) + p^{(12)}_{p-l,l} J^{(su)}_{1,2,3,4}(p-l,q)\right]
&= \sum_{l=0}^{L_u} \left[ p^{(13)}_{q,l} J^{(su)}_{1,3,2,4}(q+l,p) + p^{(13)}_{q-l,l} J^{(su)}_{1,3,2,4}(q-l,p) \right]
\, ,
\end{align}
$J^{(st)}_{a,b,c,d}(p,q)$ and $J^{(su)}_{a,b,c,d}(p,q)$ are listed in Appendix \ref{projops}.
Solving these constraint equations gives the same number of free solutions \eqref{Jamiesform} as in the degenerate case.

\subsection{Bulk interaction counting}
With canonical normalization for the kinetic term the bulk Lagrangian should be of the form
\begin{equation} \label{generallagrangian}
\mathcal{L}_{int} = \sqrt{G_N} \lambda_3 \phi^3 + G_N  \sum_{n,l,m} \lambda_{lnm}\phi \partial_{\sigma_1 \cdots \sigma_n}\partial_{\rho_1 \cdots \rho_m} \phi \partial^{\sigma_1 \cdots \sigma_n} \partial_{\tau_1 \cdots \tau_l} \phi \partial^{\rho_1 \cdots \rho_m}\partial^{\tau_1 \cdots \tau_l}\phi + \cdots,
\end{equation}
with the dimension of the $\lambda$ given by some effective field theory scale (e.g. the string scale) but not by $l_p$. All three-point interactions with derivatives reduce to four-point interactions to first order. We want to count the number of these interactions that are independent to $O(G_N^2)$ and this is equivalent to counting flat space S-matrices or monomials $s^a t^b u^c$. For the purpose of counting we will bound the spin in the $s,t,u$ channels by $L_s,L_t,L_u$ respectively, so our monomials are constrained by
\begin{equation} \label{spinbounds}
a+b \leq L_u, \qquad b+c \leq L_s, \qquad a+c \leq L_t \, .
\end{equation}
If we consider $a+b+c=n$ the spin bounds require that
\begin{equation}
 c \geq \max(0,n-L_u),\qquad  b\geq \max(0,n-L_t),\qquad a \geq \max(0,n-L_s) \, .
\end{equation}
The number of independent monomials after imposing these contraints is then the number of partitions
\begin{equation}
 a+b+c = n_{eff} = \max(0, n-\sum_i \max(0,n-L_i)) \, .
\end{equation}
However, we still have to enforce the constraint $s+t+u=4m^2$, which at the level of (\ref{generallagrangian}) comes from integrating by parts and using the equation of motion. We can use it to set $c$ to zero, which reduces our counting to partitions $a+b=n_{eff}$. There are therefore $n_{eff}+1$ independent interaction unless $n>0$ and $n_{eff}=0$. In the latter case we have 0 possibilities or we would double count $a=b=0$. We can simply encode this by moving the $+1$ inside the max, 
\begin{equation} \label{bulkcount}
 \# \textnormal{parity even solutions} = \sum_{n=0}^\infty \max(0, n+1-\sum_i \max(0,n-L_i))
\end{equation}
For the parity odd solutions in AdS$_3$ the general Lagrangian is
\begin{equation} \label{generaloddlagrangian}
\mathcal{L}_{int} = G_N  \sum_{n,l,m} \lambda_{lnm} \epsilon_{\mu\nu\kappa} \phi 
\partial^\mu \partial_{\sigma_1 \cdots \sigma_n}\partial_{\rho_1 \cdots \rho_m} \phi 
\partial^\nu \partial^{\sigma_1 \cdots \sigma_n} \partial_{\tau_1 \cdots \tau_l} \phi 
\partial^\kappa \partial^{\rho_1 \cdots \rho_m}\partial^{\tau_1 \cdots \tau_l}\phi,
\end{equation}
and as expected the counting is obtained from the previous counting by shifting the spin bounds by one in all channels, 
\begin{equation}
 \# \textnormal{parity odd solutions} = \sum_{n=0}^\infty \max(0, n+1-\sum_i \max(0,n-L_i-1))
\end{equation}
These numbers provide lower bounds on solutions to crossing, because the boundary correlators constructed from them by taking the limit are automatically conformally invariant and satisfy crossing. To see this explicitly, we worked out the partial wave expansions of several bulk amplitudes in \cite{firstpaper} and showed that they solved the crossing equations in the form \eqref{degeneratecrossing}. We have checked similar explicit solutions for the general case, but they are not further enlightening.

Although it may not be obvious at first sight, (\ref{bulkcount}) is exactly equal to (\ref{Jamiesform}) so the lower bound in this section matches the upper bound in the previous section, demonstrating that there is a local bulk theory for every boundary CFT. This is the main result of this paper.

\section{Conclusion and discussion}
\label{conclusion}
We have found further evidence for the conjecture of \cite{firstpaper} that every CFT that has a large-$N$ expansion and has parametrically large anomalous dimensions for single-trace operators with spin greater than two must have a local bulk dual. In this class of CFTs we have shown that, to order $1/N^2$, every scalar four-point function\footnote{Having considered the case in \cite{firstpaper} where the all the scalars were the same, and here where all the scalars are distinct, we don't believe that mixed cases will hold any surprises.} that is consistent with crossing has a local bulk lagrangian description. Specifically we have shown that the lower-bound for CFT four-point amplitudes found by counting local bulk interactions is saturated by an upper bound found by the consistency constraints from crossing.

While it would be nice to find an explicit map between a given CFT solution and a particular linear combination of interaction terms in the bulk, such a computation is more involved than simply counting solutions. A number of solutions were matched in \cite{firstpaper} for low $l$, but in general explicit solutions are difficult to compute both in the bulk and on the boundary.
% \footnote{For $l=0$, the CFT solutions we found for general scalars are already complicated and unenlightening. We have not included them in this paper.}

It remains interesting to extend these methods to the scattering of gravitons, as well as to conformal field theories in $d=3$. In both cases, we lack explicit expressions for the conformal partial waves, or other methods to use in their absence. The expansion in conformal partial waves seems to obscure the correspondence in solutions beneath difficult integral expressions. Finding a cleaner formalism that makes this matching transparent would be welcomed, and work continues in this direction \cite{joaoinprog}.

\subsection*{Acknowledgments}
We wish to thank J. Penedones and J. Polchinski for invaluable guidance and discussions.

\begin{appendix}

\section{Hypergeometric identities} \label{identities}

We make use of the following identities for hypergeometric functions in analytically continuing the generic amplitudes:
\begin{eqnarray}
_2F_1(a,b,c,z) &=& \frac{\Gamma(c)\Gamma(a+b-c)}{\Gamma(a)\Gamma(b)}(1-z)^{c-a-b} \, {_2F_1}(c-a,c-b,c-a-b+1,1-z) \nonumber \\
               & & + \frac{\Gamma(c)\Gamma(c-a-b)}{\Gamma(c-a)\Gamma(c-b)}  \, {_2F_1}(a,b,a+b-c+1,1-z) \nonumber \\
_2F_1(a,b,c,\tfrac{1}{z}) &=& \frac{\Gamma(c)\Gamma(b-a)}{\Gamma(b)\Gamma(c-a)}(-z)^{a} \, {_2F_1}(a,a-c+1,a-b+1,{z}) \nonumber \\
               & & + \frac{\Gamma(c)\Gamma(a-b)}{\Gamma(a)\Gamma(c-b)}  (-z)^{b} \, {_2F_1}(b,b-c+1,b-a+1,{z})
\, .
\label{idents}
\end{eqnarray}

\section{Projection operators} \label{projops}

Any degree two differential operator of the general form
\begin{equation}
\mathcal{D} = G(Z) \partial_z^2 + H(z) \partial_z
\,
\end{equation}
can be rewritten in the form
\begin{equation}
\frac{1}{h(z)}\partial_z (h(z)G(z) \partial_z) 
\, .
\end{equation}
Such an operator is self-adjoint with respect to the inner product
\begin{equation}
(F_1(z) , F_2(z) ) = \oint ( F_1(z) \cdot F_2(z)) h(z) dz \, 
\, 
\end{equation} 
over an arbitrary closed contour. This inner-product defines a projection operator on eigenfunctions of $\mathcal{D}$ provided we can choose a contour such that is non-vanishing on identical eigenfunctions. 

Using the hypergeometric equation
\begin{equation}
z(1-z)\partial_z^2 F(z) + (c -(a+b+1)z)\partial_z F(z) = a b F(z)
\,
\end{equation}
with solution $F(z) = \,_2 F_1 (a,b,c;z)$, we can construct a corresponding differential equation
\begin{eqnarray}
\mathcal{D} &=& z^2(1-z)\partial_z^2  + ((c-2a)z -(b-a+1)z^2)\partial_z \nonumber \\
           &=& (z-1)^{c-a-b}z^{2a-c+2} \partial_z \left( (z-1)^{a+b-c+1}z^{c-2a} \partial_z \right)
\, ,
\end{eqnarray}
which has eigenfunctions $V(a,b,c;z) = z^a  \,_2 F_1 (a,b,c;z)$ with eigenvalues
\begin{equation}
\mathcal{D} V = a(c-a-1) V
\, .
\end{equation}
These are eigenfunctions for the same differential operator for fixed $c-a-b$ and fixed $c-2a$. This is the case relevant to this paper where we consider arguments of the form $a = a_0 + f(n,l)$, $b = b_0 + f(n,l)$ and $c = c_0 + 2 f(n,l)$.

The naive inner-proudct vanishes on identical eigenfunctions $V$ for a contour about $z=0$. However, we have another set of solutions to the differential equation near $z=0$ given by $\tilde V(a,b,c;z) = z^{1+a -c}\, _2F_1\left(1+a -c,1+b -c,2-c;z \right)$ with identical eigenvalues:
\begin{equation}
\mathcal{D} \tilde V = a(c-a-1) \tilde V
\, .
\end{equation}
These give a projection operator
\begin{eqnarray}
\mathcal{P}_{n',l'} V(a_0+n,b_0+n,c_0+2n;z) &= \frac{1}{2\pi i } \oint& V(a_0+n,b_0+n,c_0+n;z) \nonumber\\
                                            &  &\cdot \tilde{V}(a_0+n',b_0+n',c_0+2n';z) (z-1)^{a_0+b_0-c_0}z^{c_0-2a_0-2} dz \nonumber \\
                                                         &=& \delta_{n',n}
\, . 
\end{eqnarray}

The projection operators, when acting on the non-orthogonal hypergeometric functions on the opposite side of the crossing equations generate coefficient functions given by:
\begin{eqnarray}
J^{(st)}_{a,b,c,d}(p,q)  &=& \frac{\Gamma\left(\Delta_a+\Delta_b + 2 p \right)\Gamma\left(\tfrac{\Delta_a-\Delta_b-\Delta_c+\Delta_d}{2} \right)}{\Gamma\left(\Delta_a + p \right)\Gamma\left( \tfrac{\Delta_a+\Delta_b-\Delta_c+\Delta_d}{2} + p \right)}\oint \bigg[ \frac{1}{2 \pi i z} (-1)^{p+q+1} \frac{(1-z)^{\tfrac{\Delta_a+\Delta_b-\Delta_c-\Delta_d}{2}+p}}{z^q} \, \nonumber\\
                  & &  _{2}F_{1}\left(\Delta_b + p,\tfrac{\Delta_a+\Delta_b+\Delta_c-\Delta_d}{2} + p,\tfrac{-\Delta_a+\Delta_b+\Delta_c-\Delta_d}{2} + 1 ,z \right) \nonumber\\
                  & &  _{2}F_{1}\left(1-\Delta_c-q,1-\tfrac{-\Delta_a+\Delta_b+\Delta_c+\Delta_d}{2}-q,2-\Delta_b-\Delta_c-2q ,z\right)\bigg] 
\end{eqnarray}
and
\begin{eqnarray}
J^{(su)}_{a,b,c,d}(p,q)  &=& \frac{\Gamma\left(\Delta_a+\Delta_b + 2 p \right)\Gamma\left(\tfrac{-\Delta_a-\Delta_b+\Delta_c+\Delta_d}{2} \right)}{\Gamma\left(\Delta_b + p \right)\Gamma\left( \tfrac{\Delta_a+\Delta_b-\Delta_c+\Delta_d}{2} + p \right)}\oint \bigg[ \frac{1}{2 \pi i z} (-1)^{q+1} \frac{(1-z)^{\tfrac{-\Delta_a+\Delta_b+\Delta_c-\Delta_d}{2}}}{z^q} \, \nonumber\\
                  & &  _{2}F_{1}\left(\tfrac{\Delta_a+\Delta_b+\Delta_c-\Delta_d}{2} + p,1-p-\tfrac{\Delta_a+\Delta_b-\Delta_c+\Delta_d}{2},\tfrac{\Delta_a-\Delta_b+\Delta_c-\Delta_d}{2} + 1 ,z \right) \nonumber\\
                  & &  _{2}F_{1}\left(1-\Delta_a-q,1-\tfrac{\Delta_a+\Delta_b-\Delta_c+\Delta_d}{2}-q,2-\Delta_a-\Delta_c-2q ,z\right)\bigg] 
\, .
\end{eqnarray}

\end{appendix}

% THINGS APPEAR IN THE ORDER YOU PUT THEM HERE

\end{document}